# A New Method of Obtaining High Enrichment of Metallic Single-Walled Carbon Nanotubes

Rakesh Voggu[†], A. Govindaraj[†,‡] and C. N. R. Rao[†,‡,*]

[†] *Chemistry and Physics of Materials Unit and CSIR Unit of Excellence in Chemistry, Jawaharlal Nehru Centre for Advanced Scientific Research, Jakkur P.O., Bangalore -560 064, India.*

[‡] *Solid State and Structural Chemistry Unit, Indian Institute of Science, Bangalore- 560012, India.*

**Abstract:** Arc-discharge with carbon electrodes containing Ni and $Y_2O_3$ in the presence of $Fe(CO)_5$ vapors have been investigated. It is found that unlike the cathode deposit, the web deposited in the chamber contains mainly metallic single-walled carbon nanotubes (SWNTs) as evidenced from Raman and optical absorption spectroscopy. The percentage of metallic SWNTs increases with the concentration of $Fe(CO)_5$. Under optimal conditions, the web contains over 90% of metallic SWNTs. This method enables the synthesis of nearly pure metallic SWNTs on a large scale.

(**Key words:** Carbon nanotubes, single-walled carbon nanotubes, metallic carbon nanotube)

- For correspondence: cnrrao@jncasr.ac.in, *Fax: (+91)80-22082766*

# Introduction

Single-walled carbon nanotubes (SWNTs) exhibit fascinating electronic, chemical and mechanical properties with a wide range of applications.[1,2] They can be semiconducting, semimetallic or metallic electronic types depending on the geometrical structure and have several potential applications. For example, metallic SWNTs can function as leads in nanoscale circuits and conductive additives in composites,[3-5] while the semiconducting ones can be used to design field effect transistors.[6] SWNTs can be produced by several different methods such as laser vaporization[7], arc discharge[8], high-pressure CO decomposition[9], and chemical vapor deposition[10]. Typically grown SWNTs are mixtures of metallic and semiconducting nanotubes, which limit their applicability. Recently, physical methods such as dielectrophoresis[11] and ultrcentrifugation[12, 13] as well as chemical methods including covalent and noncovalent functionalization[14] have been employed to separate metallic and semiconducting SWNTs. Selective destruction of one type of nanotubes in a mixture is also carried out by using hydrogen peroxide[15] or laser irradiation[16,17] None of these approaches, allows bulk scale production with high selectivity. An important aspect of nanotube synthesis is to develop strategies that provide narrow distributions of SWNTs of a specific electronic variety. There are reports on the diameter control of SWNTs by controlling the temperature[18], pressure[19], carbon feedstock[20] catalyst particle size[21] or catalyst material[22,23]. Methods to obtain a specific electronic type of SWNTs is of greater interest. Li *et al.*[24] and Qu *et al.*[25] have recently reported selective growth of semiconducting nanotubes using plasma-enhanced chemical vapor deposition. Enriched metallic SWNTs (up to 65%) seem to be produced by the pyrolysis of monohydroxy alcohols.[26] A method for the synthesis of pure or nearly pure metallic SWNTs has not yet been reported. Post-synthetic separations do not allow generation of bulk quantities of metallic SWNTs. We have carried out a systematic study of the effect of iron pentacarbonyl vapor on the nature of SWNTs produced in the arc discharge process. We find that the arc discharge with graphite rods containing the Ni/$Y_2O_3$ catalyst, in the presence of iron pentacarbonyl yields nanotube deposits on the walls of the discharge chamber, primarily containing metallic SWNTs( $\geq$ 90%). The procedure is simple and provides an useful method to preferentially prepare metallic SWNTs.

**Results and discussion**

The web on the walls of the discharge chamber as well as the deposit from the cathode region (Figure 1) were purified by the literature procedure.[33] In Figure 2 we show typical TEM and FESEM image of SWNTs after purification. We have examined the SWNTs collected from the web on the walls of the discharge chamber and from the cathode region ( Figure 1) by optical absorption spectroscopy as well as Raman spectroscopy, these techniques specially electronic spectroscopy, being ideally suited to distinguish the metallic and semiconducting species. the spectra of the different nanotubes samples were recorded after purification[32].

In Figure 3, we show the background-subtracted optical absorption spectra of the purified SWNTs samples prepared in the absence and presence of $Fe(CO)_5$. The sample prepared in the absence of $Fe(CO)_5$ shows three bands, a band centered around 750 nm corresponding to the metallic nanotubes ($M_{11}$) and the bands around 1040 nm ($S_{22}$) and 1880 nm ($S_{11}$) due to the semiconducting species as can be seen from Figure 3(a). The optical absorption spectra from the samples prepared in the presence of $Fe(CO)_5$ mainly show the band due to metallic species ($M_{11}$) while the bands due to the semiconducting species nearly disappear as clearly seen from the Figure 3(b). This band cannot be from any other species, since the sample had been subjected to purification and did not contain any other chemical species other than the nanotubes. It may be noted that the presence of 750 nm band in the optical absorption spectrum is taken as unequivocal evidence for the presence of metallic SWNTs.[34] Furthermore, the intensity of the $M_{11}$ band increases with the flow rate of $Fe(CO)_5$. From the integrated areas of the bands due to the metallic and semiconducting species in the optical spectrum, we estimate the relative percentage of the metallic species to be around 26 % and 94 % respectively in the absence and presence of $Fe(CO)_5$.

In order to confirm that the SWNTs from the walls of the discharge chamber obtained in the presence of $Fe(CO)_5$ was primarily metallic, we have carried out detailed Raman studies. SWNTs prepared by arc discharge exhibit bands due to the radial breathing mode (RBM) in the Raman spectra in 100-350 cm-1 region and the G-band in the 1500-1600 cm$^{-1}$ (G-band) region.[27,28] The G band of the semiconducting tubes consists of two peaks at around 1570 cm$^{-1}$ (radial) and at 1590 cm$^{-1}$ (longitudinal). The G-band of the metallic tubes shows bands around 1585 cm$^{-1}$ (radial) and at 1540 cm$^{-1}$ (longitudinal), the latter broadened into a Breit-Wagner Fano (BWF) line shape caused by strong coupling in the density of states.[29, 30] In Figure 4 we compare the Raman G-bands of SWNTs collected from the walls and the cathode regions in the absence and in the presence of $Fe(CO)_5$. The SWNTs collected from the cathode region show minor differences. This is not the case with SWNTs collected from the walls of arc chamber. The SWNTs prepared in the presence of $Fe(CO)_5$ show a marked increase in the metallic feature in the G-band. In fact, almost the entire band can be ascribed to the metallic species.

In Figure 5, we compare the RBM bands of SWNTs collected from web and cathode regions. The SWNTs prepared in the absence of $Fe(CO)_5$ show two RBM bands at 145 and 160 cm$^{-1}$ when excited with 633 nm laser radiation. Based on the revised Katauras plot, we assign the band at 145 cm$^{-1}$ to the semiconducting species (S) and the band at 160 cm$^{-1}$ to the metallic species (M).[31] Accordingly, excitation by a 514 nm laser exclusively showed only the band corresponding to the semiconducting species at 145 cm$^{-1}$.$^{32}$ The sample prepared in presence of Fe(CO)$_5$ and collected from the web exhibits an enhanced intensity of the metallic RBM band at 160 cm$^{-1}$ relative to the semiconducting band along with the evolution of two new bands at higher frequencies. The higher frequency bands which are absent in the Raman spectrum obtained with 514 nm laser excitation are attributed to the metallic nanotubes. The samples collected from cathode regions do not show much variation in the RBM bands due to the presence of Fe(CO)$_5$. In Figure 6 we show the variation of the G and RBM bands, of the SWNTs (from web region) obtained at different flow rates of Fe(CO)$_5$. We readily see the increase in the metallic features with the increase in the flow rate.

We have also compared the sheet resistance of the SWNTs from the walls of the arc chamber obtained in the presence of Fe(CO)$_5$ with those obtained in the absence of Fe(CO)$_5$. The sheet resistance is subsequently lower in the former.

**Conclusions**

In conclusion, a new method of obtaining high enrichment of the metallic SWNTs has been discovered. The method involves the arc evaporation of graphite electrodes containing the Ni+Y$_2$O$_3$ catalyst in the presence of Fe(CO)$_5$ vapor. Although we cannot comment on the mechanism of formation of metallic nanotubes in the presence of Fe(CO)$_5$ vapor, the method itself is new and simple. It can be of value in the practical use of SWNTs.

**Experimental section**

The details of the method used by us to generate SWNTs by arc discharge are as follows. Arc discharge was carried out between the two graphite electrodes in an arc discharge chamber under a helium atmosphere at 600 mbar pressure using a DC power source with a current of ~100 A and a voltage of 38 V.[8] The cathode was a pure graphite rod and the consumable anode was a composite graphite rod (6 mm diameter, 60 mm long) containing a mixture of $Y_2O_3$ (1 at.%) and Ni (4.2 at.%) catalyst. The discharge `was maintained by continuously translating the anode to keep a constant distance (~3 mm) from the cathode. The composite graphite rod was consumed during the arcing process and the cobweb like structures deposited in the chamber contained SWNTs. To prepare nearly pure metallic single walled carbon nanotubes, the arc chamber was initially filled with helium (500 mbar), and helium bubbled through iron pentacarbonyl, $Fe(CO)_5$, was then passed through the chamber at a flow rate of 200 sccm until the pressure reached to 600 mbar. Arc discharge experiments were carried out under the conditions mentioned earlier, with a continuous flow of the He+$Fe(CO)_5$ mixture, at different helium flow rates ranging from 100 to 400 sccm. Further increase in the flow rate of helium through $Fe(CO)_5$ did not favor nanotube formation. Although we cannot quantify the yield of the SWNTs with respect to $Fe(CO)_5$, we find that under an optimum concentration of $Fe(CO)_5$, the yield of SWNTs is generally greater than in the absence of $Fe(CO)_5$. SWNTs were collected from the cathode region and from the web hanging at the sides of arc discharge chamber as shown in the Figure 1. The samples collected from web and cathode region were separately subjected to purification by acid treatment followed by hydrogen treatment to remove metal and amorphous carbon impurities (Figure 2).[33]

The purified nanotubes were examined by field emission scanning electron microscopy (FESEM), transition electron microscopy (TEM), Raman and electronic spectroscopy. Raman and electronic spectroscopy give a clear indication of the proportion of the metallic and semiconducting nanotubes.

**Figure captions**

**Figure 1:** Schematic showing formation of SWNTs in the different regions of the arc-discharge chamber.

**Figrue 2:** Typcial (a) transmission electron microscopy and (b) field emission scanning electron microscopy images of metallic SWNTs after purification

**Figure 3:** G-bands in the Raman spectra of SWNTs collected from (a) the web and (b) the cathode regions (i) with the catalyst Ni+ $Y_2O_3$ alone and (ii) with the Ni+ $Y_2O_3$ catalyst in the presence of $Fe(CO)_5$ vapor .

**Figure 4: :** RBM bands in the Raman spectra of SWNTs collected from (a) the web and (b) the athode (i) with the catalyst Ni+ $Y_2O_3$ alone and (ii) with the Ni+ $Y_2O_3$ catalyst in the presence of $Fe(CO)_5$ vapor.

**Figure 5:** G and RBM bands in the Raman spectra of SWNTs (collected from the web) obtained at different $Fe(CO)_5$ flow rates.

**Figure 6:** Optical absorption spectra of SWNTs samples (collected from the web region) (a) obtained with Ni+ $Y_2O_3$ catalyst alone and (b) with Ni+ $Y_2O_3$ in the presence of $Fe(CO)_5$.

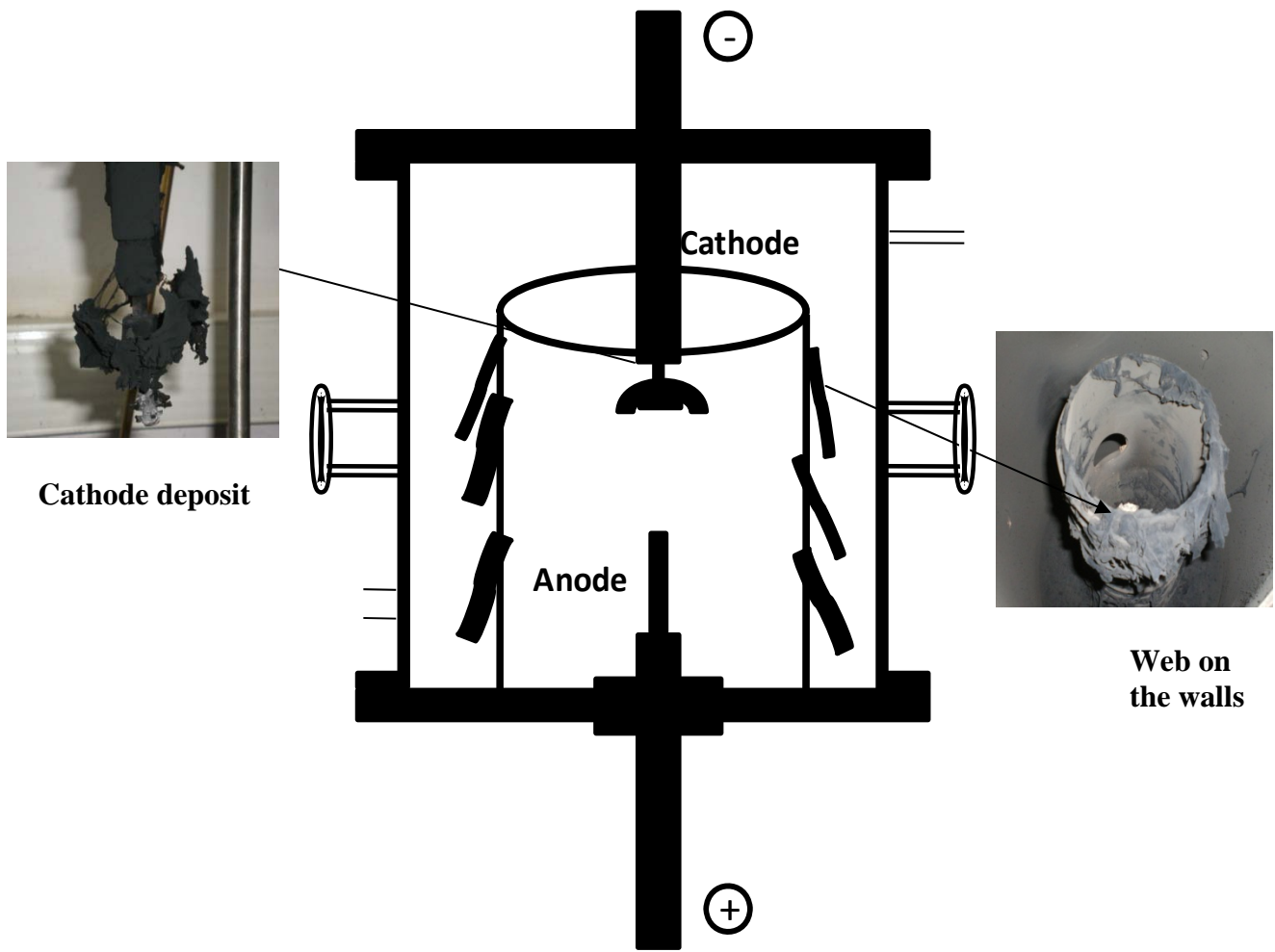

**Figure 1**

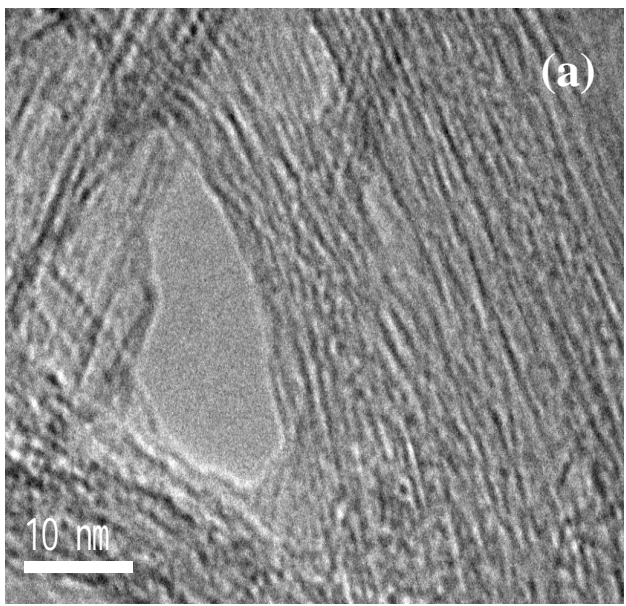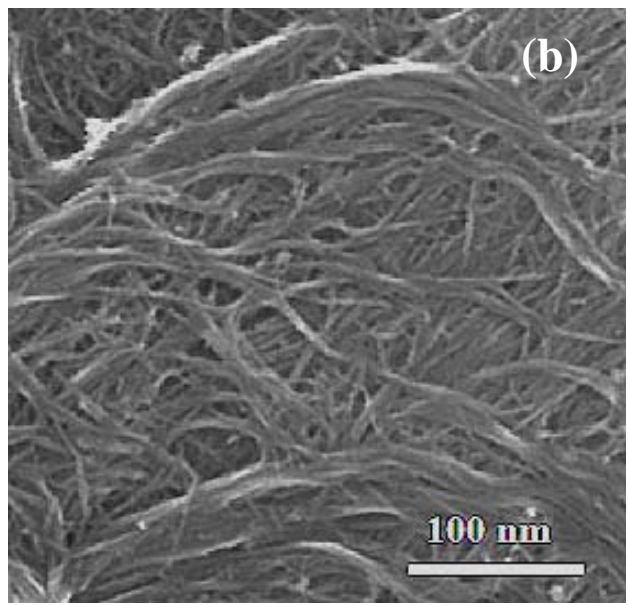

**Figure 2**

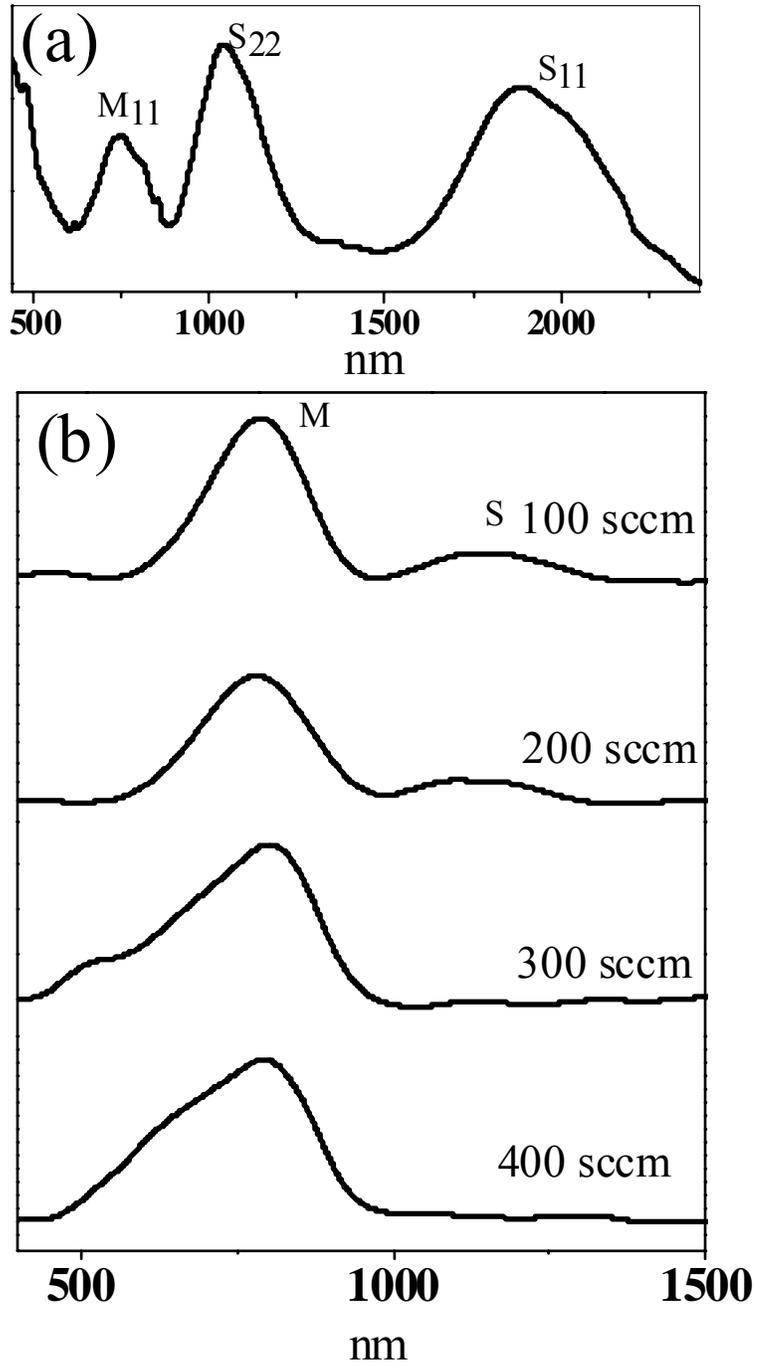

**Figure 3**

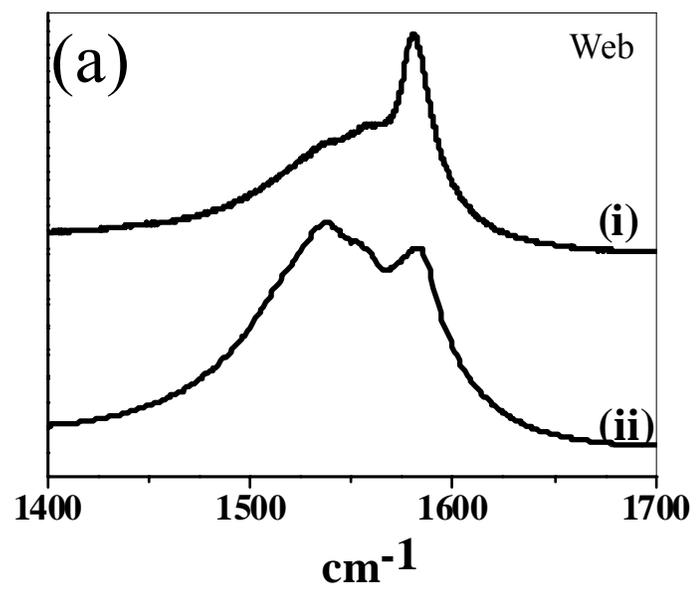

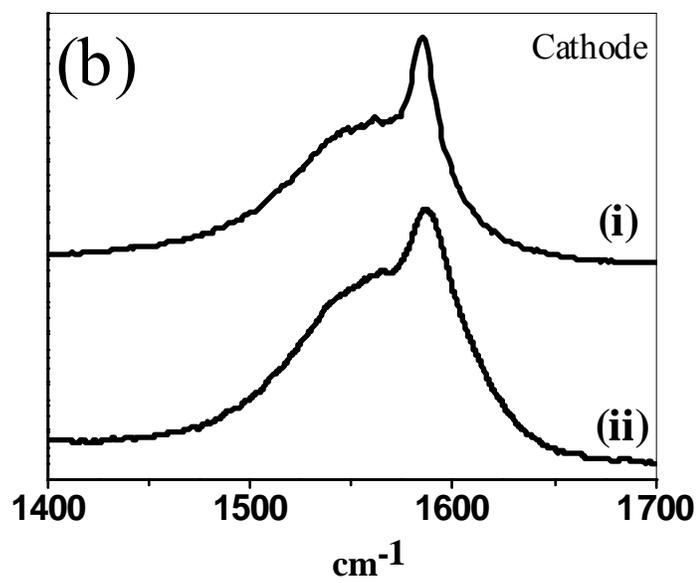

**Figure 4**

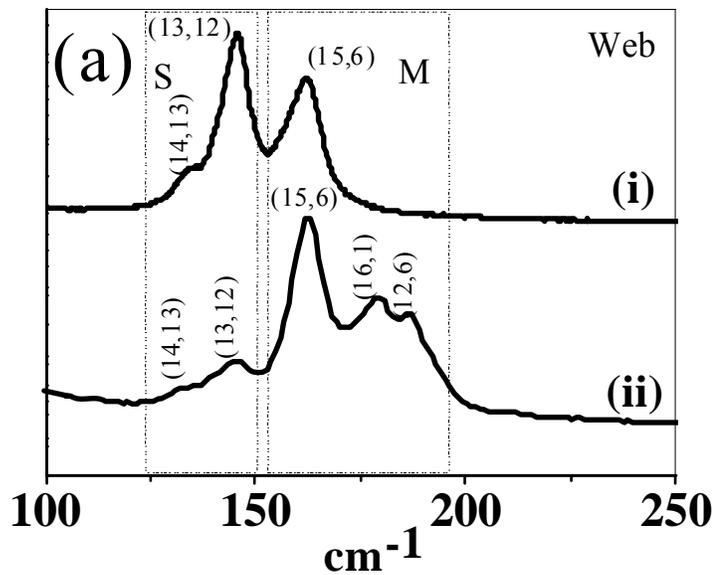
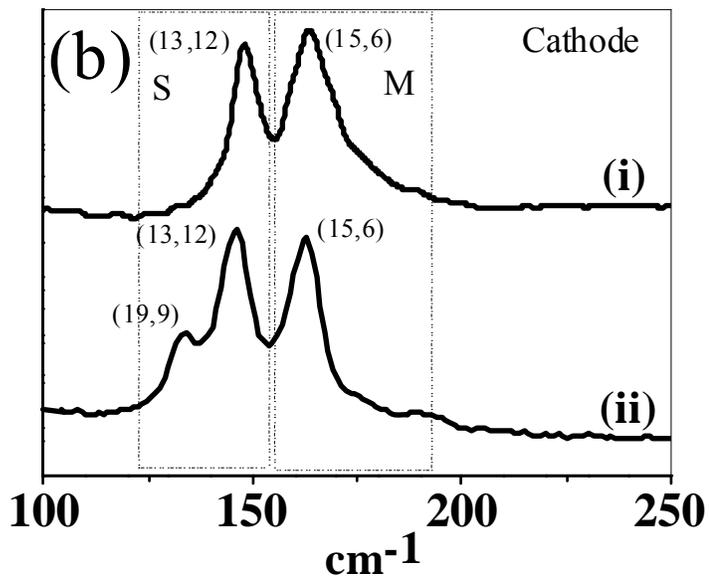

**Figure 5**

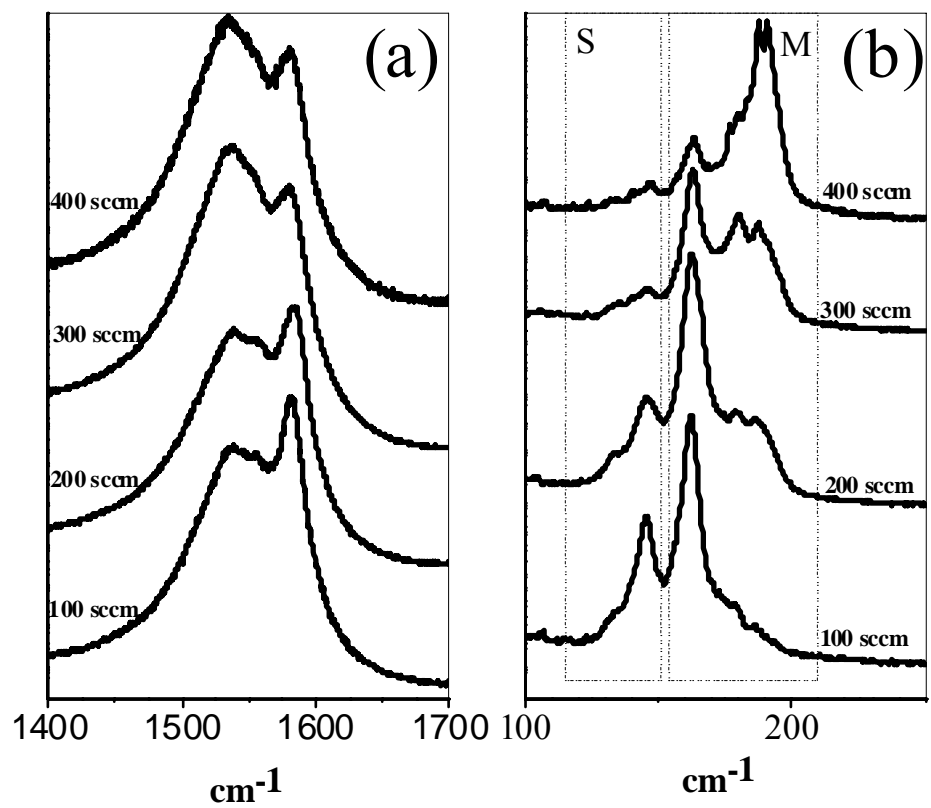

**Figure 6**